# A Randomized Concurrent Algorithm for Disjoint Set Union [*]

Siddhartha V. Jayanti [†] and Robert E. Tarjan [‡]

**Abstract**

The disjoint set union problem is a basic problem in data structures with a wide variety of applications. We extend a known efficient sequential algorithm for this problem to obtain a simple and efficient concurrent wait-free algorithm running on an asynchronous parallel random access machine (APRAM). Crucial to our result is the use of randomization. Under a certain independence assumption, for a problem instance in which there are $n$ elements, $m$ operations, and $p$ processes, our algorithm does $\Theta\left(m\left(\alpha\left(n, \frac{m}{np}\right) + \log\left(\frac{np}{m} + 1\right)\right)\right)$ expected work, where the expectation is over the random choices made by the algorithm and $\alpha$ is a functional inverse of Ackermann's function. In addition, each operation takes $O(\log n)$ steps with high probability. Our algorithm is significantly simpler and more efficient than previous algorithms proposed by Anderson and Woll. Under our independence assumption, our algorithm achieves almost-linear speed-up for applications in which all or most of the processes can be kept busy.

---

# 1 Introduction

The *disjoint set union problem*, sometimes called the *union-find problem*, requires maintaining a collection of disjoint sets under on-line union operations. Applications include storage allocation in compilers [LA02], finding minimum spanning trees [SW11], maintaining connected components in a graph under edge insertions, testing percolation [SW11], finding dominators in flow graphs [FGMT14], and computing strongly connected components in directed graphs [Blo15, BLvdP16]. The classical compressed-tree data structure solves the problem sequentially in almost-constant amortized and logarithmic worst-case time per operation. In some applications the problem instances can be enormous. A notable example is model checking [Wol02], which requires computing strongly connected components of implicitly defined, potentially huge directed graphs. A concurrent set union algorithm may well significantly improve performance in such an application, as the work of Bloemen et al. [Blo15, BLvdP16] suggests.

To our knowledge the only previous work on concurrent algorithms for set union is that of Anderson and Woll [AW91]. (Another paper [MP10] addresses the more restricted problem of doing a number of set unions simultaneously, using a distributed memory.) Anderson and Woll's computer model is an asynchronous parallel random-access machine (APRAM) [CZ89, Gib89], in which a number of processes, each with a local memory, share a common memory that contains the data structure. Concurrent reading is allowed, but not concurrent writing. There is no synchronization among the processes: any process can run arbitrarily slowly as compared to any other. The main correctness criterion is *linearizability*, the requirement that each operation can be viewed as being done atomically at some point during its execution [HW90]. An important goal in this model is to avoid having one process wait for another to finish. A concurrent algorithm in which each process can finish its operation no matter what other processes are doing is *wait-free* [Her91]. A useful primitive in building wait-free algorithms is $\text{CAS}(x, y, z)$, which tests whether $x = y$ and if so assigns $x \leftarrow z$ and returns true; if not, it merely returns false. This operation is atomic: if the test is true, no other process can change $x$ after the test but before $x$ is changed to $z$. *Total work*, the total number of primitive steps done by all processes, is the natural efficiency metric in this model.

Anderson and Woll study concurrent APRAM algorithms for disjoint set union built using CAS, with each set operation done by one process. They propose a concurrent wait-free version of a sequential algorithm, namely *linking by rank* with *path-halving* [TvL84]. They claim a total work bound of $\Theta(m(\alpha(m, 0) + p))$ for $m$ operations executed by $p$ processes. Here $\alpha$ is a functional inverse of Ackermann's function (defined in Section 2), constant for all practical purposes. The bound for the corresponding sequential (one-process) algorithm is $\Theta(m\alpha(n, m/n))$ where $n$ is the number of nodes. Thus, in a situation in which $p$ processes of equal speed are kept active concurrently throughout an execution, the total time taken by Anderson and Woll's multiprocessor algorithm to do $m$ operations is asymptotically the same as the total time taken by a sequential algorithm to do the same $m$ operations, up to a factor of $\alpha$. We conclude that Anderson and Woll's algorithm has insignificant *speed-up*. Also, to obtain their result they add a level of indirection to the data structure and add an extra computation to the union operation, both of which complicate the algorithm. Furthermore, their justification for their claimed work bound is flawed, because they completely ignore interactions among processes doing halving on intersecting paths. Even assuming their work bound (and their more-complicated results based on it) are correct, there remains the question of whether there is a simple concurrent algorithm with linear or close-to-linear speedup. Indeed, they leave this question as an open problem. We answer it in the affirmative, subject to a certain independence assumption.

Crucial to our result is randomization, specifically the use of randomized linking [GKLT14]. We develop a wait-free concurrent algorithm with an expected total work bound of

$$\Theta\left(m\left(\alpha\left(n,\frac{m}{np}\right)+\log\left(\frac{np}{m}+1\right)\right)\right)$$

and a high-probability bound of $O(\log n)$ steps for each set operation. Our computation model is that of Anderson and Woll. Our result significantly simplifies and improves theirs.

The remainder of our paper consists of six sections. Section 2 reviews efficient sequential algorithms for disjoint set union. Section 3 develops our concurrent algorithm, and Sections 4 and 5 analyze it. Section 6 discusses some variants of the algorithm, and Section 7 contains some final remarks.

## 2 Sequential Set Union using Compressed Trees

The *disjoint set union problem*, sometimes called the *union-find problem*, requires the maintenance of a collection of disjoint sets under the operation of set union. There are various versions of the problem, but we shall study the following one: Given $n$ elements, each initially in a singleton set, perform a sequence of on-line operations of the following two kinds:

SAMESET($x, y$): If elements $x$ and $y$ are in the same set, return true; otherwise, return false.

UNITE($x, y$): Given elements $x$ and $y$, if $x$ and $y$ are in different sets, unite these sets.

The classical sequential solution [Tar75, TvL84] to this problem is to represent each set by a *compressed tree*. This is a rooted tree whose nodes are the elements of the set, with each node $x$ storing a pointer $x.parent$ to its parent if it has one or to itself if it is a root. Operations SAMESET and UNITE use an auxiliary function FIND($x$) that traverses the path of parents from $x$ to the root of the tree containing $x$ and returns the root. Operation SAMESET($x, y$) returns true if FIND($x$) = FIND($y$), false otherwise. Operation UNITE($x, y$) tests whether FIND($x$) = FIND($y$) and, if not, makes FIND($x$) the *parent* of FIND($y$).

The path of nodes visited by a FIND is its *find path*. Optionally, a FIND can *compact* its find path by replacing one or more parents by nodes farther along the path. Compaction affects only efficiency, not correctness. Three classical compaction methods are *compression*, *splitting*, and *halving*. Compression replaces the parent of every node on the find path by the last node on the path (the root returned by the FIND); splitting replaces the parent of every node on the path by its grandparent; halving replaces the parent of every other node on the find path by its grandparent, starting with the first node on the path. Compression requires two passes over the find path, splitting or halving only one.

The step in UNITE that makes FIND($x$) the parent of FIND($y$) is called *linking* FIND($y$) to FIND($x$). Optionally, UNITE can choose to link FIND($x$) to FIND($y$) instead of linking FIND($y$) to FIND($x$). This choice affects only efficiency, not correctness. One classical choice is *linking by size* [Tar75], which maintains the size of each root, defined to be the number of nodes in its tree, and links the root of smaller size to the root of larger size, breaking a tie arbitrarily. A related choice is *linking by rank* [TvL84], which maintains a rank for each root, initially zero, and links the root of smaller rank to the root of larger rank; in case of a tie, it breaks the tie by increasing the rank of one of the roots by one. Both linking by size and linking by rank are deterministic; each requires maintaining a value (size or rank respectively) with each root. A more-recently studied method is *randomized linking* [GKLT14], which chooses a fixed total order of the elements uniformly at random, and links the smaller root with respect to the total order to the larger.

Any of the three compaction methods can be combined with any of the three linking methods, giving a total of nine different algorithms. Each of these has a time bound of $O(m\alpha(n, m/n))$ for $m$ operations [TvL84, GKLT14], worst-case for linking by size or rank, expected for randomized linking. Here $\alpha$ is a functional inverse of Ackermann's function, defined as follows: Let $A_k(j)$ (Ackermann's function) be the function on non-negative integers defined recursively by $A_0(j) = j + 1$, $A_k(0) = A_{k-1}(1)$ for $k > 0$, and $A_k(j) = A_{k-1}(A_k(j-1))$ for $k > 0$ and $j > 0$. For $n$ a non-negative integer and $d$ a non-negative real number, $\alpha(n, d) = \min\{i > 0 | A_i(\lfloor d \rfloor) > n\}$.

## 3 Wait-free concurrency via randomized linking

Our goal is to extend one or more of the sequential algorithms for disjoint set union to allow several set operations to be done at the same time, by separate processes. To obtain the wait-free property using CAS, every update of the data structure must produce a legal state. This makes it hard or impossible to use linking by size or rank, since both methods need to atomically update the parent of one node and the size or rank of another. Anderson and Woll implement a method related to linking by rank, using indirection in an attempt to handle the simultaneous updating problem, but they still have to handle rank ties in the data structure, which results in both an unsatisfactory work bound and complications in the implementation. Randomized linking only changes one value at a time, however, making it the method of choice for us.

Each node $x$ in our data structure has two fields, $x.parent$ and $x.id$. The forest structure is formed by the parent pointers, while the linking order is determined by the node $id$'s, fixed initially and chosen uniformly at random. If $x$ and $y$ are nodes, we use $x < y$ as shorthand for $x.id < y.id$. Initially, each node $x$ is in a one-node tree, i.e. $x.parent = x$. Anderson and Woll used one level of indirection to ensure that both the parent and the rank could be compared and updated atomically in a single CAS. We are able to do away with this indirection, since our $id$ fields are unchanged over the duration of the algorithm, and thus only parent fields need to be compared and updated atomically.

We base our concurrent implementations of SAMESET and UNITE on those of Anderson and Woll. Our implementation of SAMESET is essentially the same as theirs; our implementation of UNITE is simpler in that it does not need to do simultaneous updates and avoids indirection. In the pseudocode below, linearization points are marked by asterisks. Specifically, in each procedure the last step with an asterisk to be executed is the linearization point of the procedure. When an asterisk in a procedure is on a FIND, the linearization point of the procedure is the linearization point of the last call of the FIND.

---
**Algorithm 1** : Return the root of the tree containing node $x$.
---
1: **procedure** FIND($x$)
2:     $u \leftarrow x$
3:     **while** $(u \neq u.parent)$* **do**
4:         $u \leftarrow u.parent$
5:     **return** $u$

**Algorithm 2** : Return true if and only if $x$ and $y$ are in the same set.

1: **procedure** SAMESET($x, y$)
2:     $u \leftarrow x$
3:     $v \leftarrow y$
4:     **while** true **do**
5:         $u \leftarrow$ FIND($u$)
6:         $v \leftarrow$ FIND($v$)*
7:         **if** $u = v$ **then return** true
8:         **else if** $u = u.parent$ **then return** false

---

**Algorithm 3** If $x$ and $y$ are in different sets, unite these sets.

1: **procedure** UNITE($x, y$)
2:     $u \leftarrow x$
3:     $v \leftarrow y$
4:     **while** true **do**
5:         $u \leftarrow$ FIND($u$)
6:         $v \leftarrow$ FIND($v$)
7:         **if** $u = v$* **then return**
8:         **else if** $u < v$ **then**
9:             **if** CAS($u.parent, u, v$)* **then return**
10:        **else if** CAS($v.parent, v, u$)* **then return**

---

In SAMESET and UNITE, the linearization points are at the last execution of a step with an asterisk. These implementations rely on two key observations of Anderson and Woll: once $u = v$, nodes $x$ and $y$ are in the same set now and in the future; if $u < v$ and $u$ is a root, then $x$ and $y$ are for the moment in different sets. The implementations are more complicated than the sequential implementations described in Section 2 because they must handle the possibility that, even though $u$ is a root immediately after it is updated, $u$ can become a non-root by the time $v$ is next updated, by having another process change $u.parent$. When this happens, both $u$ and $v$ must be updated again.

Algorithm 1 does FIND without compaction. We can improve overall efficiency by doing some form of compaction. Anderson and Woll gave a concurrent version of halving. We develop concurrent versions of splitting instead. Unlike the sequential setting, in which it is unknown whether there are operation sequences on which splitting is asymptotically faster than halving or vice-versa, in the concurrent setting two processes doing halving in lockstep can simulate one process doing splitting, as the following example shows: consider a path of nodes $1, 2, \ldots, k$. Consider two processes that start at nodes 1 and 2 and do halving in lockstep. After both halvings, node $i + 2$ will be the parent of node $i$ for all $i$. This is the same as if one process had done splitting starting at node 1. One can duplicate this construction to show that, for any execution of a sequence of $m$ set operations by $p$ processes using splitting, there is an execution of a sequence of $2m$ set operations by $2p$ processes using halving, such that the latter does as many pointer updates during compaction as the former. We conclude that halving is not superior to splitting in the concurrent setting, so we choose to implement the latter.

We present two concurrent versions of splitting, *one-try* and *two-try*, Algorithms 4 and 5 respectively. One-try splitting tries once to change the parent of a node to its grandparent, after which it moves on to the next node on the find path: the update, done by a CAS, may fail, because some other process may update the parent first. Two-try splitting tries each parent update twice. We are able to derive a somewhat better efficiency bound for the latter method.

---

**Algorithm 4** Find with one-try splitting.

1: **procedure** FIND($x$)
2:     $u \leftarrow x$
3:     **while** true **do**
4:         $v \leftarrow u.parent$
5:         $w \leftarrow v.parent^*$
6:         **if** $v = w$ **then return** $v$
7:         **else** CAS($u.p, v, w$)
8:         $u \leftarrow v$

---

**Algorithm 5** Find with two-try splitting.

1: **procedure** FIND($x$)
2:     $u \leftarrow x$
3:     **while** true **do**
4:         **do twice**
5:             $v \leftarrow u.parent$
6:             $w \leftarrow v.parent^*$
7:             **if** $v = w$ **then return** $v$
8:             **else** CAS($u.parent, v, w$)
9:         $u \leftarrow v$

---

This gives us three different concurrent set union algorithms, depending on the choice of FIND method. We analyze these methods in the next section. In the remainder of this section, we sketch a proof of correctness of all three methods.

We define the *union forest* to be the forest constructed by the links done in UNITEs, ignoring all parent changes done by FINDs.

**Lemma 3.1.** *Each link in* UNITE *makes a root the child of a node in another tree. Each parent update in* FIND *replaces a parent by one of the parent's proper ancestors in the union forest. If $x$ is not a root, $x < x$.parent; if $x$ is a root, $x = x$.parent.*

*Proof.* The lemma follows by induction on the number of parent changes. CAS is used to perform links, guaranteeing that an attempt to link $u$ to $v$ by a UNITE cannot succeed unless $u$ is a root ($u.parent = u$). Furthermore, UNITE verifies that $u.id < v.id$ before attempting the link, preventing cycles from being created by concurrent link attempts. Observe that when UNITE links $u$ to $v$, $v$ may no longer be a root, even though $v$ was a root when it was last updated, since in the meantime another process could have made $v$ a child of another node. Symmetrically, when UNITE links $v$ to $u$, $u$ may no longer be a root. Finally, each CAS($u.parent, v, w$) done in a FIND is such that $u$, $v$, and $w$ are distinct, $v$ is or was the parent of $u$, and $w$ is or was the parent of $v$. By the induction hypothesis, $w$ is a proper ancestor of $v$ and $v$ is a proper ancestor of $u$ in the union forest. Thus $w$ is a proper ancestor of $u$ in the union forest. □

**Lemma 3.2** (Linearizability). *Consider any concurrent execution of set operations. The last executions of the steps marked by asterisks in each operation give a linearization. That is, let $\mathcal{O}$ be* *the order of operations given by the proposed linearization points. If the operations are executed sequentially in order $\mathcal{O}$, then their returned values will be the same as in the concurrent execution.*

*Proof.* The lemma follows by induction on the number of set operations in linearization order, and is similar to Anderson and Woll's correctness argument. Suppose the lemma holds for the first $k-1$ operations and consider the $k^{th}$.

- **Find:** For a FIND without compaction, the last execution of the equality test occurs when $u = u.parent$, ensuring that $u$ is a root at the time of its execution. For a FIND with compaction, the last execution of the equality test occurs when $w = v.parent = v$, ensuring that $v$ is a root at the time of its execution.

- **SameSet:** $u$ is always in the same tree as $x$ and $v$ is always in the same tree as $y$. We wish to show that SAMESET linearizes at the last assignment of $v$. If $v = u$ at this time, then the algorithm correctly returns true. If $v \neq u$ at this time, and $u$ is still a root, then $v$ and $u$ are both roots at the execution of the test and are thus in different trees. Thus the algorithm correctly returns false.

- **Unite:** If a UNITE does not do a link (it returns in line 7), $u = v$ at the linearization point, which guarantees that $x$ and $y$ are in the same tree at this time. If a UNITE does do a link (a CAS in line 9 or 10 succeeds), then $u \neq v$ and the smaller of $u$ and $v$ (in *id* order) is a root just before the CAS. It follows that $x$ and $y$ are in different trees just before the CAS, but in the same tree after it.

□

**Lemma 3.3** (Wait-Freedom). *The operations are wait-free.*

*Proof.* By Lemma 3.1, each update of $u$ in FIND replaces $u$ by a proper ancestor of $u$ in the union forest, and each pair of calls of FIND($u$) and FIND($v$) in SAMESET and UNITE replaces at least one of $u$ and $v$ by a proper ancestor in the union forest. It follows that any execution of SAMESET or UNITE finishes in $O(h+1)$ steps, where $h < n$ is the height of the union forest. □

**Theorem 3.4.** *The concurrent implementation is a correct linearizable wait-free algorithm for disjoint set union with any of the three versions of* FIND.

*Proof.* The theorem follows from Lemmas 3.1, 3.2, and 3.3. □

**Remark.** It is easy to extend the data structure to support the following additional operation:

MAKESET($x$): Given that $x$ is an element in no set, create a new set containing only $x$.

In a setting in which there is no a priori bound on the number of MAKESET operations (the size of the universe can grown without bound), it is possible for a SAMESET or UNITE operation to continue to make progress but never complete, because new elements are being added to the set or sets on which the operation is taking place. In such a setting our algorithms are *lock-free* rather than wait-free: at any time, there is at least one on-going process that is guaranteed to finish its operation, and any process can take a step at any time.

# 4  Analysis of Linking

Now we study the efficiency of the methods proposed in the previous section. We bound the total work (the total number of steps) taken by $p$ processes executing $m$ operations on $n$ elements. We assume that each operation is done by a single process, so at most $p$ operations are being done at any given time. In stating bounds we assume $n > 1$ (otherwise the problem is trivial) and $m \geq n/2$ (otherwise some elements are never found and can be ignored). We begin by analyzing linking. We extend the results obtained by Goel et al. [GKLT14] for sequential disjoint set union with randomized linking. Number the elements from 1 to $n$ consistent with the random total order. Identify elements by number. Define the *rank* $x.r$ of an element $x$ to be $\lfloor \lg n \rfloor - \lfloor \lg(n - x + 1) \rfloor$, where lg is the base-2 logarithm. Thus the rank of $n$ is $\lfloor \lg n \rfloor$, that of $n-1$ and $n-2$ is $\lfloor \lg n \rfloor - 1$ and so on. Since ranks are monotonic in node *id* (though not strictly), $x.r \leq x.p.r$ for any node $x$.

We say a UNITE *succeeds* if it does a CAS that returns true; that is, it does a link, and *fails* otherwise. We make the following independence assumption:

(∗) The random node order is independent of the linearization order of the of UNITEs.

**Lemma 4.1.** *If $v$ is a proper ancestor of $u$ in the union forest, the probability that $v.rank > u.rank$ is at least $1/2$, independent of the other proper ancestors of $u$.*

*Proof.* Since the proof relies heavily on the ideas of Goel et al. we merely sketch it. Fix the linearization order of the UNITEs and fix a node $u$. Goel et al. show that one can arrange the UNITEs that succeed into an order $\sigma(u)$ so that each one unites the set containing $u$ with a singleton set, and that, no matter what the random node order, the set of ancestors of $u$ in the union forest contains all the ancestors of $u$ in any union forest formed by doing the UNITEs in order using the same random node order. This proof extends to the concurrent setting: for a fixed random node order, a fixed linearization of the UNITEs will produce a union forest in which the ancestors of $u$ are a subset of those of $u$ in the union forest produced by executing the UNITEs sequentially in order $\sigma(u)$. Furthermore, Goel et al. [GKLT14] show that if the UNITEs are executed sequentially in order $\sigma(u)$, then each proper ancestor of $u$ has probability at least $\frac{1}{2}$ of having rank higher than that of $u$, independent of the rest. The lemma follows, since the rank of a node is never less than the rank of its *parent*.  □

**Corollary 4.1.1.** *For any node $x$, the expected number of ancestors of $x$ of the same rank as $x$ is $O(1)$.*

*Proof.* By Lemma 4.1, the expected number of ancestors of $x$ of the same rank is at most $\sum_{i=0}^{\infty} 1/2^i = 2$.  □

**Lemma 4.2.** *Let $h \geq 1$ and let $x$ be any node. The probability that $x$ has an ancestor $y$ in the union forest of rank less than $h$ and such that the path from $x$ to $y$ contains $ch$ parent pointers is at most $1/exp((c/2 - 1)^2 h^2 / c)$.*

*Proof.* Consider the links that formed the path $P$ from $x$ to $y$ in the union forest. Call such a link *good* if it links two nodes of different ranks and *bad* otherwise. Since ranks along a path in the union forest are non-decreasing, there are fewer than $h$ bad links among the $ch$ links that formed $P$. By Lemma 4.1, each link is good with probability at least $1/2$, independent of the other links. The Lemma follows by a Chernoff bound [CL05].  □

**Corollary 4.2.1.** *With probability at least $1 - 1/n$, the height of the union forest is $O(\log n)$.*

*Proof.* Let $h = \lg n + 1$, where lg is the base-two logarithm. Choose $c$ in Lemma 4.2 so that the probability that a node $x$ has an ancestor $y$ $ch$ arcs higher in the union forest is at most $1/n^2$. Summing over all $n$ nodes, the probability that there is any path in the union forest of $ch$ arcs is at most $1/n$. Hence the height of the union forest is at most $c(\lg n + 1)$ with probability at least $1 - 1/n$. □

**Corollary 4.2.2.** *Let $h \geq 1$. For some constant $c > 0$ the expected number of nodes of rank less than $h$ and height at least $ch$ in the union forest is at most $n/exp((c/2 - 1)^2 h^2/c)$.*

*Proof.* For a node $y$ to have height at least $ch$ in the union forest, it must have a descendant $x$ $ch$ arcs deeper in the union forest. For any node $x$, the probability that it has an ancestor $y$ $ch$ arcs higher in the union forest of rank less than $h$ is at most $1/exp((c/2 - 1)^2 h^2/c)$ by Lemma 4.2. Summing over all $n$ nodes gives the corollary. □

Given an input $x$ or $y$ to a SAMESET or UNITE, we define the *find sequence* of $x$ or $y$ to be the sequence of values of $u$ or $v$, respectively, during the operation. By Lemma 3.1, each find sequence is a sequence of distinct, not necessarily consecutive ancestors in the union forest. In each pair of of FINDs done by a SAMESET or UNITE, other than the first, at least one of $u$ and $v$ is not a root. It follows that the number of steps done by SAMESET$(x, y)$ or UNITE$(x, y)$ is $O(1)$ plus a constant times the number of nodes in the find sequences of $x$ and $y$.

**Theorem 4.3.** *With any of the three versions of FIND, the probability that every operation does $O(\log n)$ steps, and hence that the total work is $O(m \log n)$, is at least $1 - 1/n$.*

*Proof.* By Corollary 4.2.2, with probability at least $1 - 1/n$, every find sequence contains $O(\log n)$ nodes. □

**Remark.** We can increase the success probability in Corollary 4.2.1 and Theorem 4.3 to $1 - 1/n^{c \log n}$ for any positive constant $c$, at the cost of increasing the constant hidden in the big-O notation.

**Remark.** Corollary 4.2.1 confirms the conjecture that the union forest in the sequential version of randomized linking is of logarithmic height with high probability, a problem left open in [GKLT14].

By Theorem 4.3, concurrent randomized linking has linear speedup with high probability over a sequential method that uses any linking rule but no compaction.

## 5 Analysis of Splitting

We now derive better work bounds for the two algorithms that combine randomized linking with concurrent splitting. Our analysis extends that of Goel et al. [GKLT14] to the concurrent setting.

We use the *potential method* of amortized analysis: we assign to each configuration of the data structure a real-valued *potential* and define the *amortized cost* of an operation to be its actual cost plus the net increase in potential it causes. The total cost of a sequence of operations is then the sum of their amortized costs plus the initial potential minus the final potential. In the analysis of set union, the potential is always non-negative; thus the total cost of a sequence of operations is at most the sum of their amortized costs plus the initial potential.

We need to extend the analysis of Goel et al. [GKLT14] to deal with one process changing a parent pointer while other processes are visiting the same node. This would seem to increase the work bound by a factor of $p$ over the sequential time bound. But we are able to reduce this to roughly a logarithmic factor by combining two ideas: we multiply the node potential of Goel et al. by a factor of $2p$, and we give positive potential only to nodes of sufficiently high rank, of which there are very few. The need to count steps that fail to change parent pointers was missed by Anderson and Woll, who claimed (falsely) that an $O(m\alpha(n, m/n))$ time bound for a sequential set union algorithm implies an $O(m\alpha(n, m/n))$ work bound for a concurrent version of the same algorithm. We begin by analyzing two-try splitting: even though it is more complicated than one-try splitting, its analysis is simpler.

Let $d$ be a non-negative value, to be chosen later. We define the *index function* $b(i, k)$ for integers $i$ and $k$ by $b(i, k) = \min\{j \geq 0 | A(i, j) > k\}$. We define the *level function* $a(k, j)$ for non-negative integers $k$ and $j$ by $a(k, j) = \min(\{\alpha(k, d) + 1\} \cup \{i \leq \alpha(k, d) | A(i, b(i, k)) > j\})$. For each node $x$, we define a *level* $x.a$, *index* $x.b$, and *count* $x.c$ as follows: $x.a = a(x.r, x.parent.r)$; $x.b = b(x.a - 1, x.parent.r)$ if $x.a > 0$, 0 otherwise; $x.c = x.a(x.r + 2) + x.b$.

Goel et al. [GKLT14] proved the following for sequential splitting. It is straightforward to verify that their proof extends to one-try and two-try splitting as well:

(i) The level of a node is non-negative, non-decreasing, and at most $\alpha(n, d) + 1$.

(ii) The count of a node is non-negative and non-decreasing.

(iii) If the level of a node increases, its count increases by at least as much.

(iv) The level of a node is 0 if and only if it has the same rank as its parent.

(v) If a node has level 0 and its parent changes, it decreases in potential.

(vi) Let $u$ be any node such that $u$, $u.parent$, and $u.parent.parent$ are distinct. If $1 \leq u.a \leq u.parent.a$, then changing the parent of $u$ to $u.parent.parent$ or any higher node increases $u.c$ by at least 1. If $u.a < u.parent.a$, then changing the parent of $u$ to $u.parent.parent$ or any higher node increases $u.a$ to at least $u.parent.a$.

We use these six properties to bound the total work done by our algorithm if FINDs use two-try splitting.

**Theorem 5.1.** *If* FIND*s are done with two-try splitting, the expected total work done by the concurrent set union algorithm is* $O\left(m\left(\alpha\left(n, \frac{m}{np}\right) + \log\left(\frac{np}{m} + 1\right)\right)\right)$, *assuming the random node order is independent of the linearization order of the* UNITE*s.*

*Proof.* Let $d = \frac{m}{np}$. We consider two cases. The first is $d \geq 1$. We define the potential of a node $x$ to be $2p$ times the number of proper ancestors of $x$ of the same rank as $x$, plus $2p$ times $\max\{0, (\alpha(x.r, d) + 1) \times (x.r + 2) + d + 1 - x.c\}$, and the potential of the data structure to be the sum of the node potentials. This is $2p$ times the potential as defined in Goel et al. [GKLT14]. By their results and Corollary 4.1.1, the potential of each node is non-negative and non-increasing, and the expected total node potential is $O(n(d + 1)p) = O(m)$.

The total work is $O(m)$ plus at most a constant times the number of occurrences of nodes on FIND paths. The number of FINDs is at most $2m + 2np \leq 4m$ since each SAMESET or UNITE starts by doing two FINDs, and each link, of which there are at most $n - 1$, can give rise to $2p$ additional FINDs, two by each process. We define the cost of a FIND to be one less than the number of executions of the while loop. The number of steps done by a FIND is $O(1)$ plus a constant times the number of time it executes the while loop, so defining the cost in this way undercounts the number of steps by a constant factor and an additive $O(m)$ term.

When a process starts a FIND, say FIND($x$), we give the process a potential of $\alpha(n, d) + 1$ to complete the FIND. When a node decreases in potential, we assign $1/p$ of the decrease to each of the $p$ processes. We prove that the cost of a FIND is at most the potential assigned to it during the find, from which it follows that the total work done by all processes on all $m$ set operations is $O(m\alpha(n,d))$, giving the theorem for the case $d \geq 1$.

To prove that the potential assigned to a process pays for its loop iterations, we prove something a little stronger:

(*) Each time the value of the variable $u$ in FIND($x$) is updated, the process executing the FIND has at least $u.a$ units of potential.

We prove (*) by induction on the number of assignments to $u$. Suppose a process, say process 1, begins executing FIND($x$). Since $u.a \leq \alpha(n,d) + 1$ for any node $u$ at any time, the potential assigned to process 1 when it begins executing FIND($x$) is enough to satisfy (*) when process 1 first assigns a value to $u$. Suppose (*) holds immediately after some assignment to $u$, say at time $t_1$, and consider the next assignment to $u$, say at time $t_2$. Between $t_1$ and $t_2$, process 1 executes the steps in the while loop once. Let $u_1$ and $u_2$ be the values of $u$ at $t_1$ and $t_2$, and let $u_1.a(t)$ and $u_2.a(t)$ be the levels of $u_1$ and $u_2$ at time $t$. We need to show that, between $t_1$ and $t_2$, process 1 accrues at least $u_2.a(t_2) - u_1.a(t_1) + 1$ units of potential: this is enough to guarantee that (*) holds at $t_2$ after process 1 spends one unit of potential to do an iteration of the while loop. To show that process 1 accrues the needed potential, we establish the following crucial inequality:

(**) $u_1.a(t_2) \geq u_2.a(t_1)$

To prove (**), we consider two cases. If the second CAS done by process 1 between $t_1$ and $t_2$ succeeds, then after this CAS the level of $u_1$ is at least $u_2.a(t_1)$ by (i) and (vi), which implies (**) by (i). Suppose the second CAS in the iteration of the loop fails. This happens because a CAS done by some other process, say process 2, succeeds in updating $u_1.parent$ between the first CAS and the second CAS done by process 1. Furthermore, for the CAS done by process 2 to succeed, process 2 must have set its variable $v$ equal to $u_2$ after $t_1$, since this must happen after the first update to $u_1.parent$ between $t_1$ and $t_2$. Such a first update occurs because either the first CAS done by process 1 between $t_1$ and $t_2$ succeeds, or a CAS done by some other process updates $u_1.parent$ between the time process 1 sets its variable $v$ for the first time after $t_1$ and process 1 does its first CAS. After the successful CAS done by process 2, the level of $u_1$ is at least $u_2.a(t_1)$ by (i) and (vi), which implies (**) by (i). Thus (**) holds in either case.

We use (**) as follows. Between $t_1$ and $t_2$, process 1 accrues at least $2(u_1.a(t_2) - u_1.a(t_1) + u_2.a(t_2) - u_2.a(t_1)) = u_2.a(t_2) - u_1.a(t_1) + u_1.a(t_2) - u_1.a(t_1) + u_1.a(t_2) - u_2.a(t_1) + u_2.a(t_2) - u_2.a(t_1)$ units of potential. By (**), this is at least $u_2.a(t_2) - u_1.a(t_1)) + 1$, enough to give (*) at time $t_2$, unless $u_1.a(t_1) = u_1.a(t_2) = u_2.a(t_1) = u_2.a(t_2)$; that is, the levels of $u_1$ and $u_2$ are equal and unchanging between $t_1$ and $t_2$. But if the levels of $u_1$ and $u_2$ are equal and unchanging between $t_1$ and $t_2$, by (v) (if $u_1.a(t_1) = 0$) or (vi) (if $u_1.a(t_1) > 0$) process 1 accrues at least one unit of potential between $t_1$ and $t_2$ as a result of either its second CAS between $t_1$ and $t_2$ succeeding, or the CAS that makes its second CAS fail succeeding. We conclude that (*) holds update-by-update, hence in general by induction, giving the theorem for $d \geq 1$.

The second case is $d < 1$; that is, $m < np$. In this case we apply the argument in the first case only to nodes of high rank. We apply separate arguments to nodes of low height in the union forest and to nodes of low rank but high height in the union forest, of which there are very few by Corollary 4.2.2. Let $h = \lceil \lg(1/d) \rceil$. We call a node *high* if its rank is at least $h$, *low* if its height in the union forest is less than $4h$, and *middle* if it is neither high nor low.

The number of low nodes in any find sequence is at most $4h$, so the total number of occurrences of low nodes on find sequences is at most $4mh = 4m \lg(1/d)$. To count high nodes, we define a find to be *high* if it visits at least one high node. The high finds include all occurrences of high nodes on find sequences. There are $O(n/2^h) = O(nd) = O(m/p)$ high nodes. The number of high finds is at most two per set operation plus at most $2p$ per link of a high node to another (high) node, for a total of $O(m)$. We define the *reduced rank* of a high node to be its actual rank minus $\lceil \lg(1/d) \rceil$. If we define node potentials as in the first case, but only for high nodes and using reduced ranks in place of actual ranks, so that they start at 0, the argument in the first case shows that the expected total number of occurrences of high nodes in find sequences is the sum of the find potentials, which is $O(m)$, for a total of $O(m\alpha(n,d))$.

By Corollary 4.2.2, the expected number of middle nodes is at most $n/2^{h^2/16}$. The parent of a middle node is either middle or high. Once a middle node acquires a high parent, its parent stays high. When a middle node acquires a high parent, we charge it $p$, to account for the $p$ or fewer processes that may be visiting it at the time. If a middle node on a find path is not a root or a child of a root, its parent changes as a result of the find. By Corollary 4.1.1, the expected number of times this can happen before its parent has rank at least $h$ and hence is not middle is $O(h)$. We charge $p$ for each such change, to account for the $p$ or fewer processes that may be visiting the node at the time. The remaining possibility is that of a middle node on a find path that is a root or the child of a root. Either the node is last or next-to-last on its find sequence, or it or its parent acquires a parent. The latter can happen at most twice per middle node, each of which we charge $p$ to account for the $p$ or fewer processes that may be visiting the node at the time. The expected total charge to all middle nodes is $O\left(hpn/2^{h^2/4}\right) = O(m)$.

Combining all three of our estimates, the expected total number of occurrences of nodes on find sequences is $O(m(\alpha(n,d) + \log(1/d)))$.

□

We now present an upper bound for one-try splitting. In the conference version of our paper [JT16], we claimed that Theorem 5.1 holds for one-try splitting, but our proof, essentially the proof of Theorem 5.1, is not correct for this algorithm. The problem is that a CAS done by one process, say process 1, can fail as a result of a successful CAS done by another process, say process 2, that sets its value of $v$ *before* process 1's current iteration of the while loop. In order to make sure that process 1 receives enough potential to maintain (*), we must increase the node potentials and allow processes that do successful CAS operations to transfer their excess potential to other processes. This results in a slightly worse bound than that of Theorem 5.1.

**Theorem 5.2.** *If* FIND*s are done with one-try splitting, the expected total work done by the concurrent set union algorithm is $O\left(m\left(\alpha\left(n, \frac{m}{np^2}\right) + \log\left(\frac{np^2}{m} + 1\right)\right)\right)$ assuming the random node order is independent of the linearization order of the* UNITE*s.*

*Proof.* Since many parts of the proof are the same as those in the proof of Theorem 5.1, we just note the changes to these parts. We redefine $d = \frac{m}{np^2}$. We again consider two cases, $d \geq 1$ and $d < 1$. In the case $d \geq 1$, we give each node the same potential as in the proof of Theorem 5.1 but call it a *normal potential*. In addition, we give each node an *excess potential* equal to $p-1$ times its normal potential. That is, each node gets a potential equal to $2p^2$ times the node potential of Goel et al. [GKLT14]. The expected total node potential (normal and excess) is $O(n(d+1)p^2) = O(m)$. When the potential of a node decreases, we assign a fraction $1/p$ of the decrease in both normal and special potential to each of the processes. We discuss the further transfer of excess potential below.

As in the proof of Theorem 5.1, we prove (*) by induction on the number of assignments to $u$ by a process, say process 1, executing a FIND, from which the Theorem follows for the case $d \geq 1$.

In one-try splitting there is only one assignment to $v$ inside the while loop; $u_2$ is the value of $v$ immediately after this assignment. The proof of the induction step is the same as in the proof of Theorem 5.1 if the CAS done by process 1 succeeds. There is only one CAS inside the while loop; we obtain the proof by replacing "the second CAS" by "the CAS" throughout the proof of (*) in Theorem 5.1.

We need to expand the argument slightly if the CAS done by process 1 between $t_1$ and $t_2$ fails. Such a failure is the result of a CAS done by another process, say process 2, succeeding. In this case let $t_0$ be the last time before this successful CAS that process 2 sets its variable $u$. When process 2 does its successful CAS, it transfers to process 1 a fraction $\frac{1}{p-1}$ of the excess potential it accrues as a result of decreases in the excess potentials of $u_1$ and $u_2$ between $t_0$ and the time of its CAS. The argument that proves (**) in the proof of Theorem 5.1 shows that $u_1.a(t_2) \geq u_2.a(t_0)$.

If $t_0 \geq t_1$, (*) holds at $t_2$ by the same argument as in the proof of Theorem 5.1, using only the normal potential accrued by process 1 between $t_1$ and $t_2$. Suppose on the other hand that $t_0 < t_1$. Process 1 obtains at least $2(u_1.a(t_1) - u_1.a(t_0) + u_2.a(t_1) - u_2.a(t_0))$ units of special potential from process 2 when the CAS done by process 2 succeeds. This and the normal potential accrued by process 1 between $t_1$ and $t_2$ give process 1 at least $2(u_1.a(t_2) - u_1.a(t_0) + u_2.a(t_2) - u_2.a(t_0)) = u_2.a(t_2) - u_1.a(t_0) + u_1.a(t_2) - u_1.a(t_0) + u_1.a(t_2) - u_2.a(t_0) + u_2.a(t_2) - u_2.a(t_0)$ units of potential. This is at least $u_2.a(t_2) - u_1.a(t_1)) + 1$, enough to give (*) at time $t_2$, unless $u_1.a(t_0) = u_1.a(t_2) = u_2.a(t_0) = u_2.a(t_2)$; that is, the levels of $u_1$ and $u_2$ are equal and unchanging between $t_0$ and $t_2$. But if the levels of $u_0$ and $u_2$ are equal and unchanging between $t_0$ and $t_2$, by (v) (if $u_1.a(t_0) = 0$) or (vi) (if $u_1.a(t_0) > 0$) process 1 accrues at least one unit of potential between $t_0$ and $t_2$ as a result of the CAS of process 2 between $t_0$ and $t_2$ succeeding. This gives (*), from which the theorem follows for $d \geq 1$.

The proof for the case $d < 1$ is the same as the proof of the corresponding case in Theorem 5.1. □

We conclude this section by showing that the bound in Theorem 5.1 is tight to within a constant factor. We need the following lemma.

**Lemma 5.3.** *For any $k$, a suitable sequence of $k-1$ UNITEs whose FINDs are done by one-try or two-try splitting will build a $k$-node tree whose average node depth is $\Omega(\log k)$.*

*Proof.* The proof is constructive and is similar to the construction of binomial trees. The idea is the same as that used to build binomial trees; namely, start with singletons, unite them in pairs, unite the resulting trees in pairs, and repeat until there is one tree. The only complication is that the construction process does not have direct access to the tree roots but can only access them via FINDs, which do splitting. By choosing the parameters of the UNITEs properly, we can guarantee that the splits have little effect.

Specifically, suppose $k$ is a power of 2. Begin with $k$ singletons, each of whose nodes is the representative of its set. Repeat the following $\lg k$ times: Pair up the existing sets, unite the pairs by calling UNITE on the representatives of the two sets, and designate either of the representatives of the two old sets to be the representative of the new set. This process maintains the following invariants:

(1) After $i$ rounds of UNITEs, each tree contains $2^i$ nodes.

(2) Each representative has depth at most two.

(3) After $i$ rounds of UNITEs, the subtree of any node $x$ of depth $\delta$ contains at most $2^{i-\delta}$ nodes.

Invariant (1) is immediate. Invariant (2) follows by induction on the number of rounds of UNITEs, since if (2) holds for a representative, a FIND on it reduces its depth to at most 1, and a subsequent link can increase its depth to at most 2. Invariant (3) follows since splitting can only decrease the number of descendants of a non-root.

Now consider the sum of the depths of all $k$ nodes as UNITEs proceed. Consider a UNITE in round $i+1$ that combines two trees with representatives $x$ and $y$, respectively. The FIND on $x$ can decrease one or more node depths only if $x$ has depth 1 before the FIND and 1 after, in which case it decreases the sum of node depths by at most $2^{i-2}$ by (3). Similarly, the FIND on $y$ can decrease the sum of node depths by at most $2^{i-2}$. The link after the two FINDs increases the sum of node depths by $2^i$, for a net increase of $2^{i-1}$. The entire round increases the sum of node depths by at least $\frac{k}{4}$. We conclude that after all $\lg k$ rounds, the average node depth is at least $\frac{1}{4} \lg k$.

If $k$ is not a power of 2, we apply the argument above to the largest power of 2 no larger than $k$ (at least $\frac{k}{2}$) to form a tree $T$, form an arbitrary tree out of enough singletons to bring the total number of nodes to $k$, and unite the two trees, doing the find in $T$ on its representative. □

**Theorem 5.4.** *The bound in Theorem 5.1 is tight to within a constant factor. That is, for any $n$ and $m$ there is a set of $m$ operations on $n$ elements done by $p$ processes such that the expected work using two-try splitting FINDs is*

$$\Omega\left(m\left(\alpha\left(n, \frac{m}{np}\right) + \log\left(\frac{np}{m} + 1\right)\right)\right)$$

*Proof.* The lower bound proof is in two parts. First we derive a lower bound of $\Omega\left(m\alpha\left(n, \frac{m}{np}\right)\right)$. Next we obtain a lower bound of $\Omega\left(m \log\left(\frac{np}{m}\right)\right)$. Putting the two together yields the theorem.

1. Fredman and Saks [FS89] proved a sequential lower bound of $\Omega(m\alpha(n, m/n))$ for $m \geq n$, that holds in expectation for randomized algorithms that expose the operations UNITE and FIND. Our version of the problem only exposes SAMESET and UNITE operations. But we can do exactly the same amount of work as a FIND($x$) if we simply add an extra node $\xi$ so that there are a total of $n+1$ nodes, and then do a SAMESET($x, \xi$) instead of doing a FIND($x$). Thus, if operations are done sequentially, by the single process $p_1$ using our algorithm, there must be a sequence of $M$ operations that do $\Omega(M\alpha(n, M/n))$ work for each $M \geq n$. We consider two cases for the range of $m$:

   (a) If $m \geq np$ then $\frac{m}{p} \geq n$. Thus we can set $M = \frac{m}{p}$ and get a sequence of $\frac{m}{p}$ operations that do a total of $\Omega(\frac{m}{p}\alpha(n, \frac{m}{np}))$ work. If we do the same operations on each of the $p$ processors, and the processors run in lockstep, we get $p \cdot \frac{m}{p} = m$ operations that take $\Omega(m\alpha(n, \frac{m}{np}))$ expected work.

   (b) If $n \leq m < np$ then $\frac{m}{n} < p$, we instead consider $M = n$. This gives us a sequence of $n$ operations for a single process that take $\Omega(n\alpha(n, 1))$ expected work. If we simply assign these operations to $\frac{m}{n}$ of the processors, and the scheduler makes the corresponding operations on different processors run in lockstep, the expected work is $\Omega(\frac{m}{n} \cdot n \cdot \alpha(n, 1)) = \Omega(m\alpha(n, 1)) = \Omega(m\alpha(n, \frac{m}{np}))$ since $\frac{m}{np} < 1$.

2. In the case that the $\log\left(\frac{np}{m}\right)$ term dominates the $\alpha$ term, we know that $np > m$. Let $\delta = \frac{1}{3d} = \frac{np}{3m}$ for simplicity of notation in the lower bound example below, and note that $\delta < n$, since $n < 2m$ and $l < m$. Below is an explicit sequence of operations that take $\Omega(m \log \delta)$ expected time (we assume without loss of generality that $\delta$ divides $n$):

(a) Use one of the processes to create $\frac{n}{\delta}$ trees $T_1, \ldots, T_{\frac{n}{\delta}}$ each with expected node depth $\Omega(\log \delta)$. This can be done by Lemma 5.3 using at most $n$ UNITE operations.

(b) Choose nodes $x_1, \ldots, x_{\frac{n}{\delta}}$ randomly from the trees $T_1, \ldots, T_{\frac{n}{\delta}}$ respectively.

(c) Perform the operation SAMESET$(x_i, x_i)$ with each of the $p$ processes for each $i \in \{1, \ldots, \frac{n}{\delta}\}$. If the processes run in lockstep, then each operation takes $\Omega(\log \delta)$ work in expectation.

Note that since $m = \Omega(n)$, at least a constant fraction of the operations are SAMESET operations from part (c). It follows that the total expected work of the set of operations is:

$$\Omega\left(\frac{n}{\delta} \cdot p \cdot \log \delta\right) = \Omega(m \log \delta) = \Omega\left(m \log\left(\frac{np}{m}\right)\right)$$

□

The lower bound in Theorem 5.4 holds for the algorithm with either one-try or two-try splitting. For two-try splitting, it yields a tight bound. For one-try splitting, our upper and lower bounds are very close, but not equal. We leave open the problem of closing the gap.

## 6 Variants

Just as Goel et al. [GKLT14] did in the sequential setting, we can take advantage of the total order of nodes to terminate one of the two FINDs in both SAMESET and UNITE early, if we do the two FINDs concurrently. The idea is to take a step from whichever of the two current nodes on the two FIND paths is smaller, stopping when the smaller current node is a root. Early termination of FINDs can be done whether or not some form of compaction is used. The bounds in Sections 4 and 5 (for the appropriate form of compaction) hold if early termination is used. Algorithms 6 and 7 implement SAMESET and UNITE, respectively, with early termination and two-try splitting. We get the one-try splitting variant if we execute the statements in the "do twice"-loop only once per while-loop iteration.

---

**Algorithm 6** Same-set implemented with early-recognition.
---
 1: **procedure** SAMESET$(x, y)$
 2:     $u \leftarrow x \quad v \leftarrow y$
 3:     **while** true **do**
 4:         **if** $u = v^*$ **then return** true
 5:         **if** $v < u$ **then** SWAP$(u, v)$
 6:         **if** $u = u.parent^*$ **then return** false
 7:         **do twice**
 8:             $z \leftarrow u.parent$
 9:             $w \leftarrow z.parent$
10:             CAS$(u.parent, z, w)$
11:         $u \leftarrow z$
---

**Algorithm 7** Unite with early termination and two-try splitting.

```
 1: procedure UNITE(x, y)
 2:     u ← x    v ← y
 3:     while true do
 4:         if u = v* then return
 5:         if v < u then SWAP(u, v)
 6:         if CAS(u.parent, u, v)* then return
 7:         do twice
 8:             z ← u.parent
 9:             w ← z.parent
10:             CAS(u.parent, z, w)
11:         u ← z
```

Goel et al. [GKLT14] also analyze two additional forms of compaction: compression and splicing. Splicing does the two FINDs of a UNITE by traversing the two FIND paths concurrently, changing each parent of a node on one path to a node on the other path. Splicing during a concurrent UNITE seems dangerous, since it may splice together parts of two different trees before reaching the linearization point of the UNITE, possibly resulting in incorrect behavior. On the other hand, we conjecture that appropriate concurrent versions of compression will have the bounds of Theorems 5.1 and 5.2. We think though, that some version of splitting is the method of choice in a concurrent setting, since splitting requires only one traversal of the FIND path (compression requires two) and is purely local.

## 7  Remarks

We have designed and analyzed three wait-free concurrent algorithms for disjoint set union. Each uses randomized linking. All three methods are work-efficient, assuming that the linearization order of the UNITEs is independent of the initially chosen random node order. Our best bounds are for two-try splitting; whether one-try splitting has the bound of Theorem 5.1 is an open problem.

We have assumed that the node order is chosen before any set operations are done. If new elements can be added by the MAKESET operation (see Section 3), one can generate the node order on the fly, by assigning to each new element a random number selected uniformly from a universe large enough that the chance of a tie is sufficiently small, and adding a tie-breaking rule to prevent cycles from being formed by concurrent links. For example, if the elements are stored in a hash table, the storage location of an element can be used as its number.

Our efficiency results rely on a somewhat questionable independence assumption. In new work, we have developed several concurrent versions of linking by rank that give the bounds of Sections 4 and 5. One of them is randomized and needs no independence assumption. The other two are deterministic. We shall report on these results in a future paper.